\documentclass{PoS}

\usepackage{graphicx,wrapfig,multirow,xspace,amsmath,amssymb}

\newcommand{\sqsn}{\ensuremath{\sqrt{s_\mathrm{NN}}}\xspace}
\newcommand{\sqs}{\ensuremath{\sqrt{s}}\xspace}
\newcommand{\pT}{\ensuremath{p_\mathrm{T}}\xspace}
\newcommand{\kT}{\ensuremath{k_\mathrm{T}}\xspace}
\newcommand{\Dzero}{\ensuremath{\mathrm{D}^{0}}\xspace}
\newcommand{\Dstarp}{\ensuremath{\mathrm{D}^{*+}}\xspace}
\newcommand{\Dsp}{\ensuremath{\mathrm{D}_s^{+}}\xspace}
\newcommand{\Dp}{\ensuremath{\mathrm{D}^{+}}\xspace}

\title{Open heavy-flavour production in pp and p--Pb collisions with the ALICE experiment}

\ShortTitle{Open heavy-flavour in pp and p--Pb collisions with ALICE}

\author{\speaker{R\'obert V\'ertesi} (for the ALICE Collaboration)\\
        Wigner Research Centre for Physics RMI, Budapest, Hungary\\
        E-mail: \email{vertesi.robert@wigner.mta.hu}}

\abstract{Measurements of the heavy-flavour hadron production are a powerful tool to study the nature of strong interaction and to understand the properties of nuclear matter that is created in ultra-relativistic hadron-hadron collisions.
The excellent tracking and vertexing capabilities of the ALICE experiment, together with its particle identification systems, allow for an efficient identification and reconstruction of decays of hadrons that contain heavy quarks.
A selection of recent measurements on open heavy-flavour production from pp collisions at $\sqs=7$ TeV and p--Pb collisions at $\sqsn=5.02$ TeV, collected with the ALICE detector during the LHC Run-1 phase, are discussed in this paper.}

\FullConference{XXV International Workshop on Deep-Inelastic Scattering and Related Subjects\\
		3-7 April 2017\\
		University of Birmingham, UK}

\begin{document}

\section{Introduction}

Charm and bottom quarks are produced in hard processes, in the early stages of relativistic hadron-hadron collisions. Due to their large masses, their initial production yield is mainly unaffected by the later stages of the reaction. Therefore, they provide us with unique means for understanding several aspects of Quantum Chromodynamics, such as pure pQCD processes, energy loss and collective motion within a nuclear medium created in collision systems of various sizes, coalescence of heavy and light quarks, as well as fragmentation depending on the quark mass~\cite{Andronic:2015wma}.
Measurements of total yields and spectra in pp collisions serve as a benchmark for theory calculations as well as a reference for collisions of heavier systems. The multiplicity-dependence of the production may reveal the importance of multi-parton interactions, the interplay between the hard and soft regime of the interactions, as well as the connection between open and hidden production of heavy-flavour~\cite{Frankfurt:2010ea}. Measurements in p--Pb collisions also account for cold nuclear matter (CNM) effects, providing the baseline to quantify effects of a hot and dense medium in heavy-ion measurements. Different CNM effects such as the modification of the parton distribution functions in nuclei (nPDF) by (anti)shadowing or by gluon saturation, \kT{}-broadening by multiple soft scatterings and energy loss in the CNM, are expected to give contributions that may also depend on the choice of the rapidity window~\cite{Fujii:2013yja,Mangano:1991jk,Sharma:2009hn,Kang:2014hha}.
Correlations of D mesons and charged hadrons provide insight to charm fragmentation and may as well reveal possible collective effects in p--Pb collisions~\cite{Mangano:1991jk}.

\section{Experiment}
Hadrons containing heavy quarks are either detected directly via the reconstruction of hadronic decays, or indirectly by finding a single electron or muon that is produced via a semi-leptonic decay channel. Direct reconstruction has the obvious advantage that the whole kinematics is under control and the mother hadron can be identified unambiguously. However, reducing the combinatorial background will result in a significant loss in reconstruction efficiency. Leptons from heavy-flavour decays, on the other hand are characteristic signals, with higher branching ratios, that can be directly triggered on. However, the leptons from heavy-flavour decays will be a mixture of contributions from decays of charmed and beauty hadrons. Secondary vertex identification techniques are used both to aid identification in the direct reconstruction, as well as to statistically separate the charm and the beauty contributions of semi-leptonic decays.

ALICE (A Large Hadron Collider Experiment)~\cite{Aamodt:2008zz} carries out measurements of D-mesons (e.g.\ \Dzero, \Dp, \Dstarp and \Dsp), as well as electrons from heavy-flavour decays (HFE), in the central barrel at mid-rapidity. Charged tracks are reconstructed with the Time Projection Chamber (TPC) and the Inner Tracking System (ITS). Particles are identified by specific ionisation energy loss $\mathrm{d}E/\mathrm{d}x$ in the TPC, aided by the Time-of-Flight detector (TOF). The ITS is also used for the statistical separation of electrons from charm and beauty decays based on the distance of closest approach to the primary vertex, as well as for background reduction by secondary vertex reconstruction in the D-meson measurements. The minimum bias trigger is provided by the V0 detectors placed at forward rapidity. The Electromagnetic Calorimeter (EMCal) serves as a triggering device, and for the total energy measurement of electrons. Muons from heavy-flavuor decays (HFM) are measured in the Muon Spectrometer in the $-4<|\eta|<-2.5$ rapidity window.

\section{Production in pp collisions}

\begin{figure}
\begin{minipage}{0.48\textwidth}%
\includegraphics[width=\textwidth]{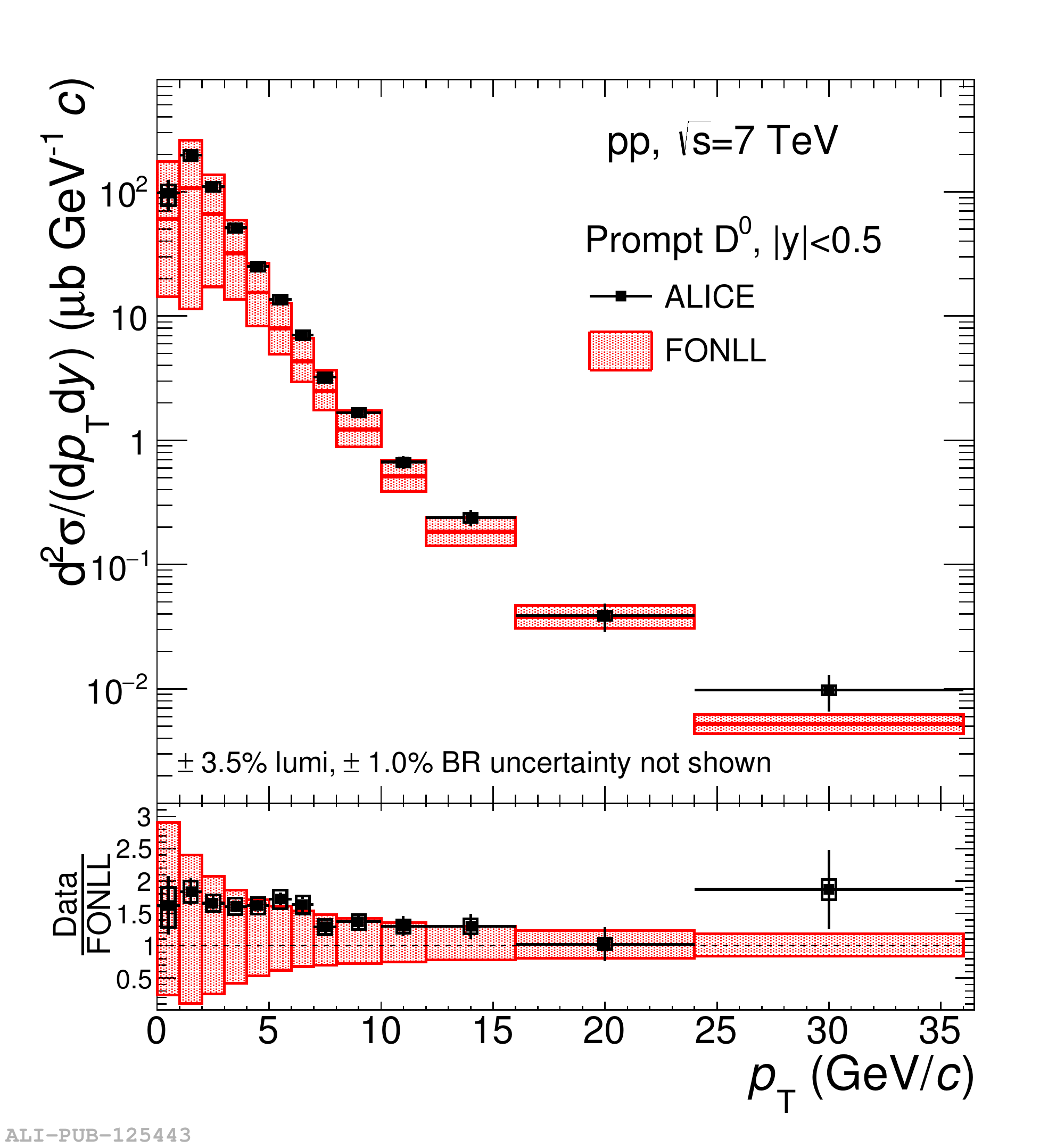}%
\caption{\Dzero \pT-differential invariant cross section at |y|<0.5 in pp collisions at $\sqs=7$ TeV~\cite{Acharya:2017jgo}, compared with FONLL calculations~\cite{Cacciari:2012ny}.}%
\label{fig:D_pp}%
\end{minipage}%
\hspace{0.04\textwidth}
\begin{minipage}{.48\textwidth}%
\includegraphics[width=\textwidth]{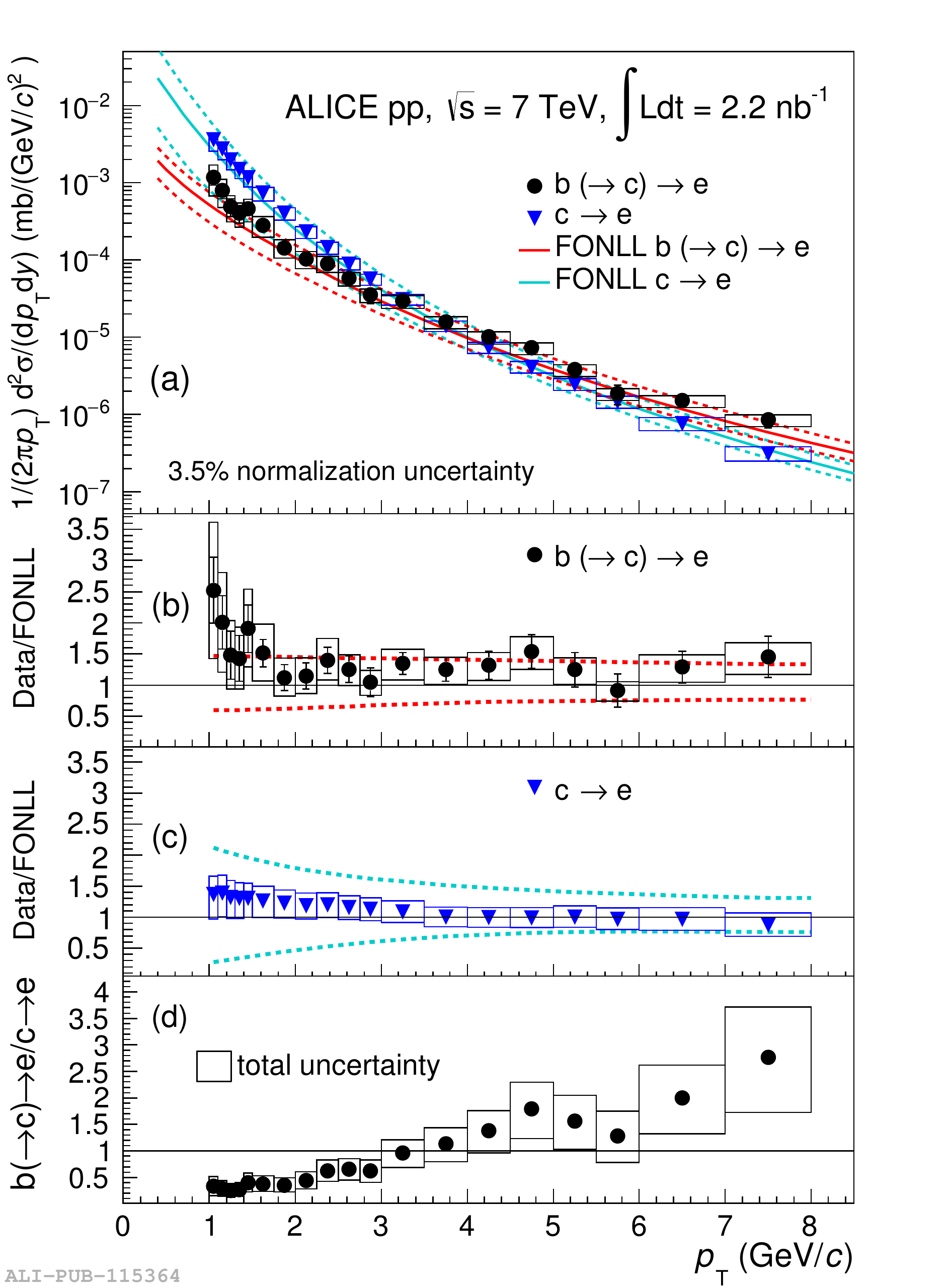}%
\caption{The \pT-differential invariant cross sections of electrons from charm and beauty hadron decays in pp collisions at $\sqs=7$ TeV~\cite{Abelev:2012sca}, compared with FONLL calculations~\cite{Cacciari:2012ny}.}%
\label{fig:HFE_pp}
\end{minipage}
\end{figure}

The total $c\bar{c}$ production cross section in pp collisions at $\sqrt{s}=7$ TeV, measured via D-meson decays~\cite{Acharya:2017jgo}, is 
$\sigma^{c\bar{c}}_{\mathrm{pp},7\ \mathrm{TeV}} = 8.08 \substack{+2.55\\-1.04} \ \mathrm{mb}$. The \Dzero{}-meson \pT{}-differential cross section at mid-rapidity is shown on Fig.~\ref{fig:D_pp} from zero transverse momentum up to $\pT=35$ GeV/$c$. FONLL pQCD calculations~\cite{Cacciari:2012ny} are consistent with these data within uncertainties.
The total $b\bar{b}$ cross section, measured using electrons from beauty decays~\cite{Abelev:2012sca}, is 
$\sigma^{b\bar{b}}_{pp,7\ \mathrm{TeV}} = 322 \substack{+74\\-78} \ \mu\mathrm{b}$. 
Figure~\ref{fig:HFE_pp} shows the \pT-differential cross sections of electrons from separated beauty and charm decays in pp collisions at $\sqrt{s}=7$ TeV at mid-rapidity. FONLL calculations~\cite{Cacciari:2012ny} describe data within uncertainties in the full momentum range.

\section{Nuclear matter effects in p--Pb collisions}

\begin{figure}
\includegraphics[width=.5\textwidth]{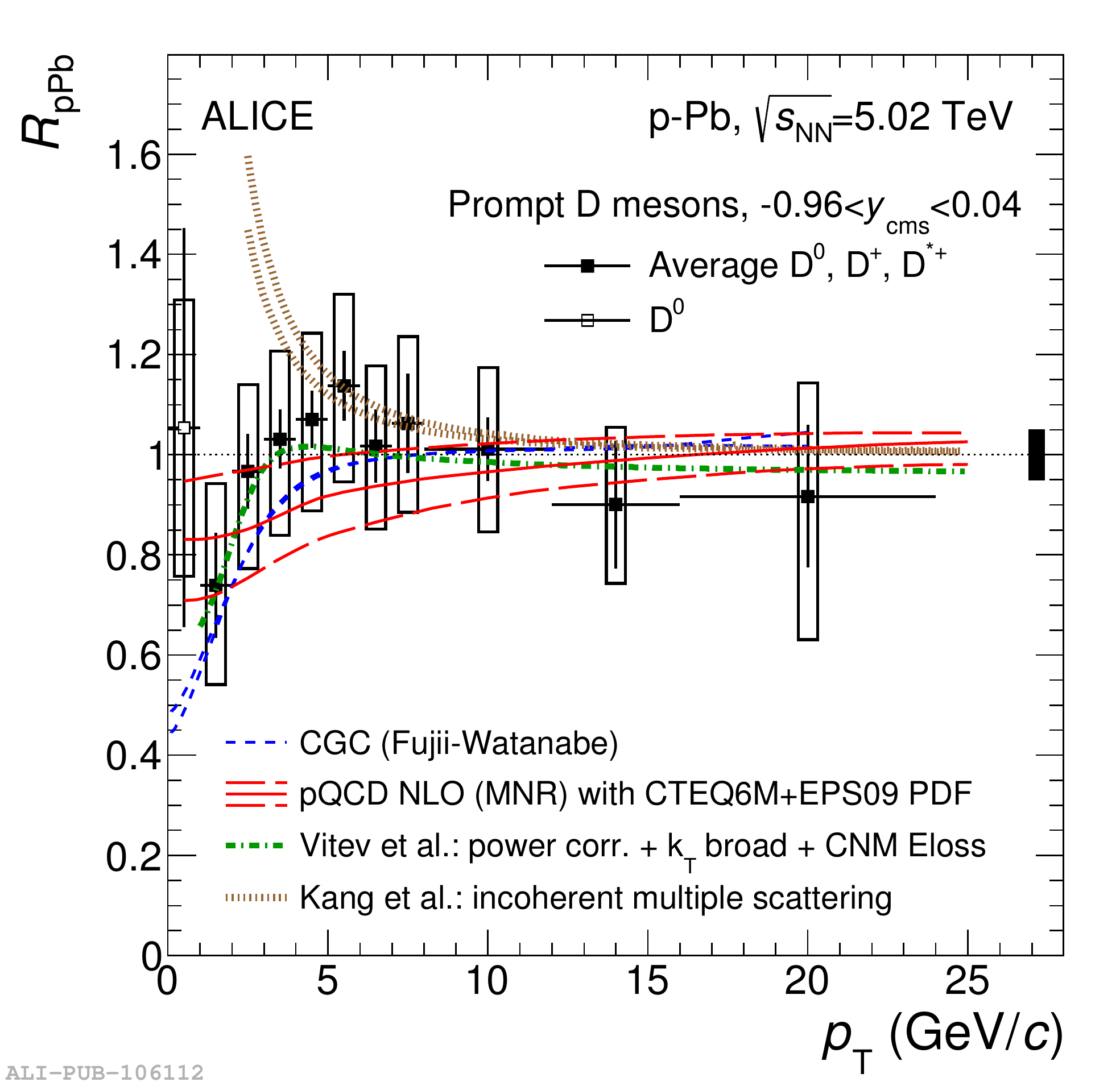}%
\includegraphics[width=.5\textwidth]{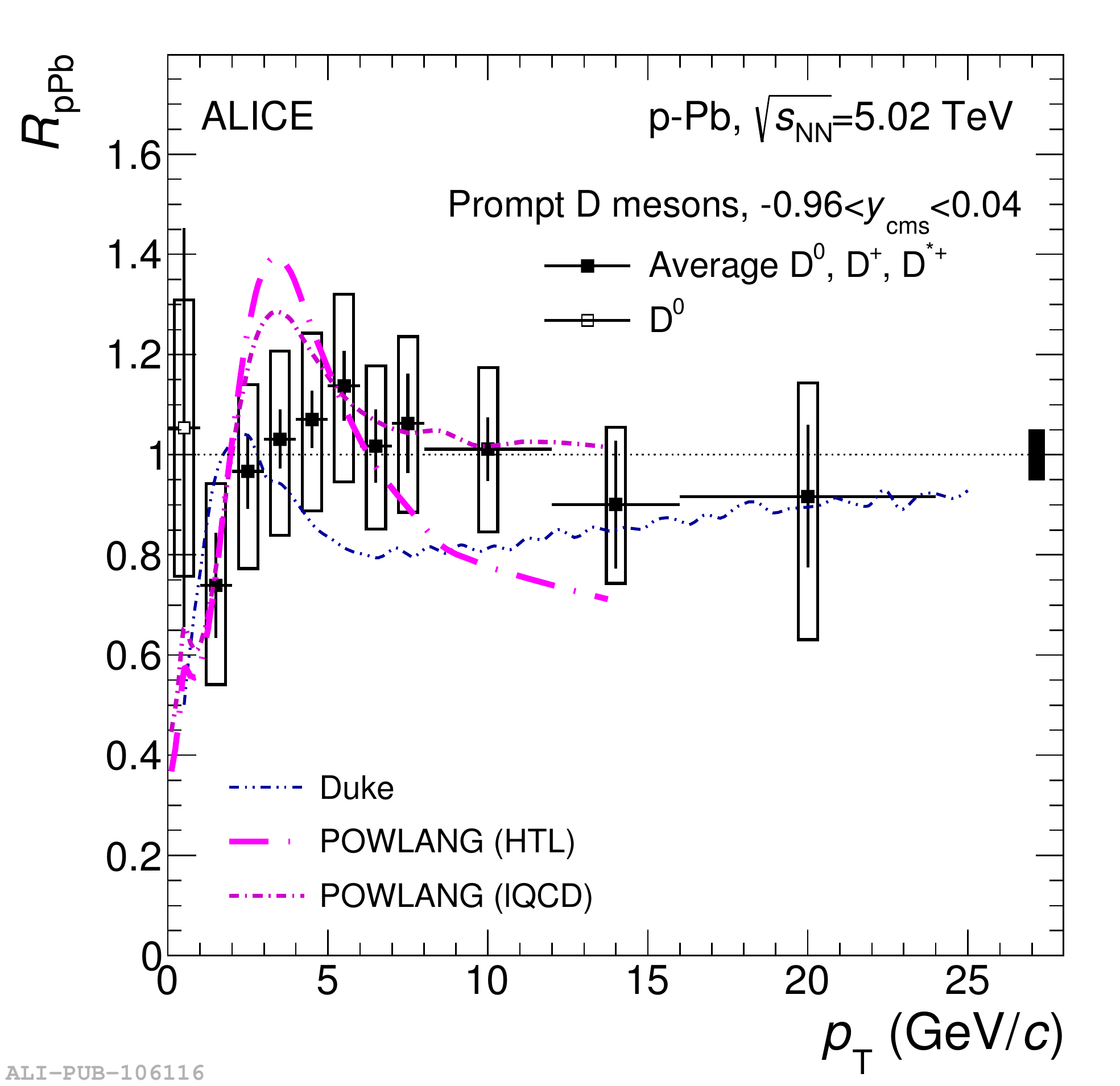}
\caption{Nuclear modification factor $R_{\rm pPb}$ of prompt D mesons in p--Pb collisions at $\sqsn=5.02$ TeV at central rapidity~\cite{Abelev:2014hha}. The data are compared models assuming CNM effects~\cite{Fujii:2013yja,Mangano:1991jk,Sharma:2009hn,Kang:2014hha} ({\it left}), as well as additional hot medium (Quark-Gluon Plasma) effects in a small volume~\cite{Nardi:2015pca} ({\it right}).}
\label{fig:D_pPb}
\end{figure}

Figure~\ref{fig:D_pPb} shows the nuclear modification of D mesons in p--Pb collisons at $\sqs=5.02$ TeV~\cite{Abelev:2014hha}. The measurements are consistent with no nuclear modification in the whole momentum range. The results are described by several theoretical calculations that include CNM effects (Fig.~\ref{fig:D_pPb} left panel)~\cite{Fujii:2013yja,Mangano:1991jk,Sharma:2009hn}. 
The low-\pT enhancement predicted by calculations with incoherent multiple scattering~\cite{Kang:2014hha} is not supported by the data.
The fact that we do not observe a strong nuclear modification in p--Pb collisions, adds experimental evidence that the modification seen in Pb--Pb collisions is caused by a hot and dense deconfined medium~\cite{Abelev:2014hha}. It is to be noted, however, that some models that assume the creation of such a medium in p--Pb collisions in a small volume are also capable of describing the data~\cite{Nardi:2015pca} (Fig.~\ref{fig:D_pPb} right panel).
 
\begin{figure}
\includegraphics[width=.5\textwidth]{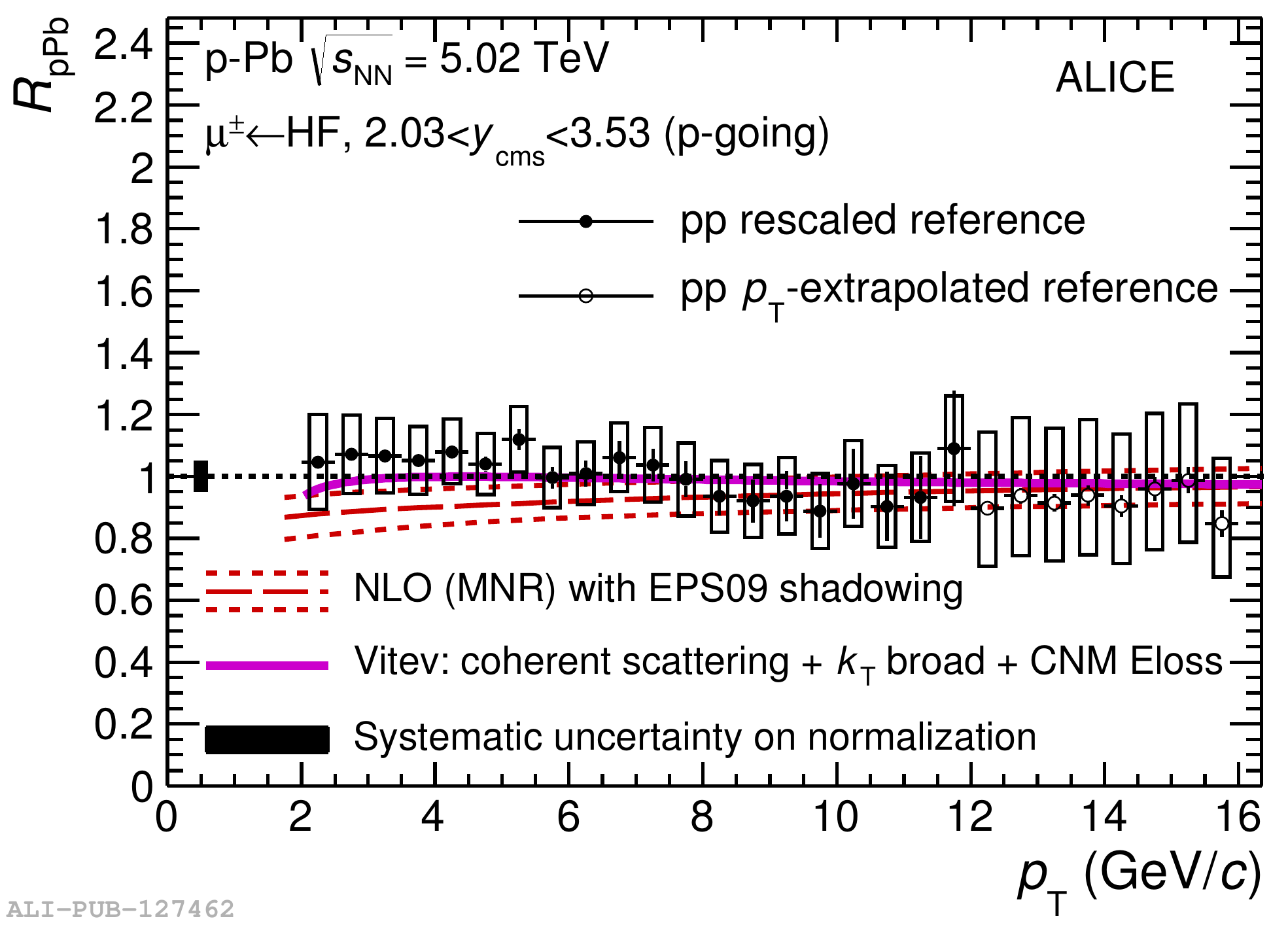}%
\includegraphics[width=.5\textwidth]{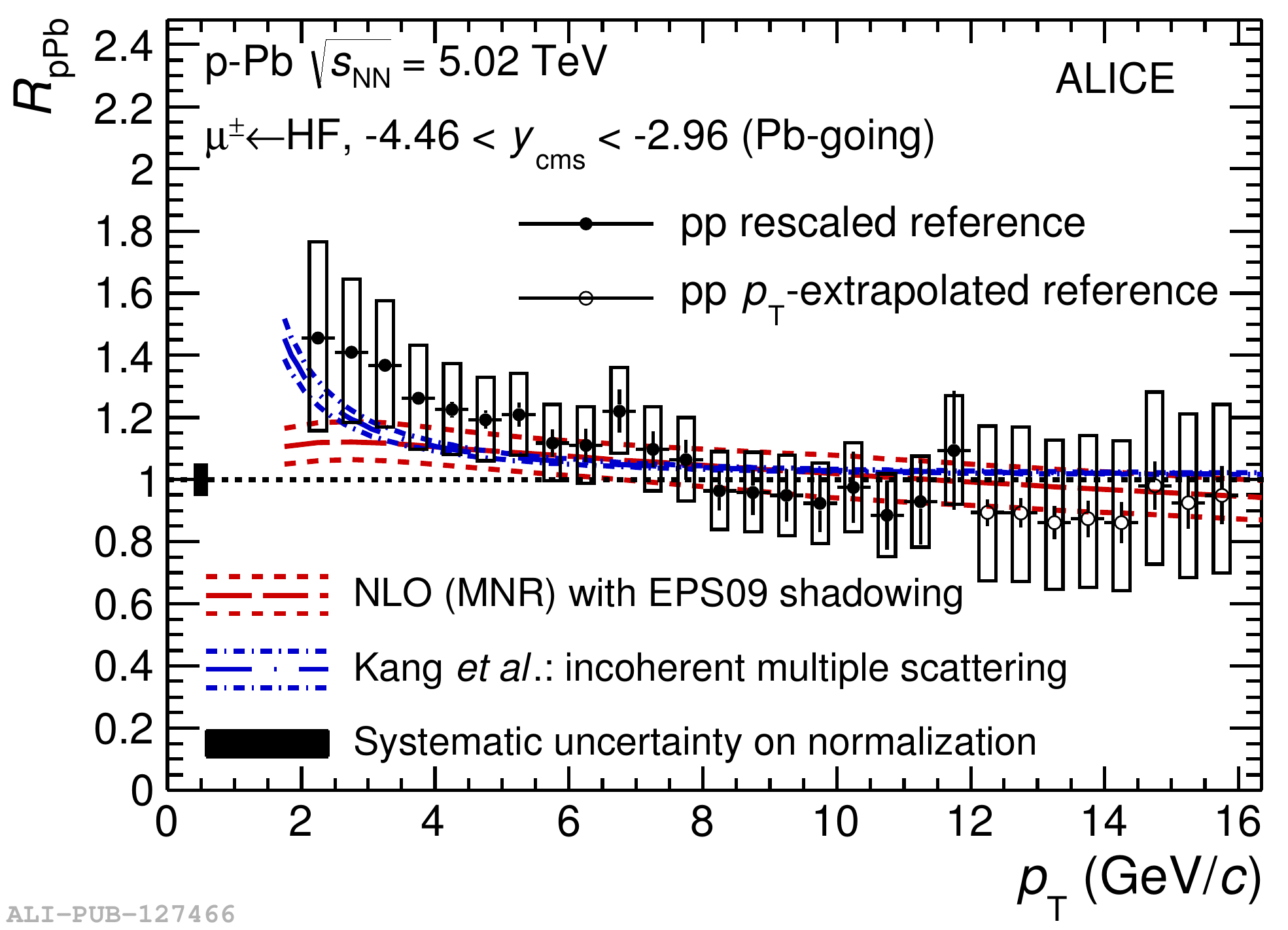}
\caption{Nuclear modification factor of muons from heavy-flavour hadron decays as a function of \pT for p--Pb collisions at $\sqsn=5.02$ TeV at forward ({\it left}) and backward rapidity ({\it right}), compared to model predictions~\cite{Mangano:1991jk,Sharma:2009hn,Kang:2014hha}.}
\label{fig:mu_pPb}
\end{figure}

The nuclear modification factor of muons is shown on Fig.~\ref{fig:mu_pPb} in the forward ($2.03<y_\mathrm{cms}<3.53$) and backward ($-4.46<y_\mathrm{cms}<2.96$) rapidity ranges~\cite{Acharya:2017hdv}. The corresponding Bjorken-$x$ values of gluons probing the Pb nucleus in a LO pair production process, are  in the order of $10^{-5}$ and $10^{-1}$, respectively~\cite{Acharya:2017hdv}. While the modification in the forward, shadowing region is consistent with unity, some enhancement is indicated at lower \pT values in the backward, anti-shadowing region. The backward and forward nuclear modification factors are simultaneously reproduced within uncertainties by calculations including nuclear modification of the PDFs~\cite{Mangano:1991jk}, while other models containing shadowing, \kT{}-broadening and CNM energy loss~\cite{Sharma:2009hn}, or incoherent multiple scattering~\cite{Kang:2014hha}, reproduce either forward or backward results.

\section{Multiplicity dependence in heavy-flavour production}

\begin{figure}
\begin{minipage}{0.48\textwidth}%
\includegraphics[width=\textwidth]{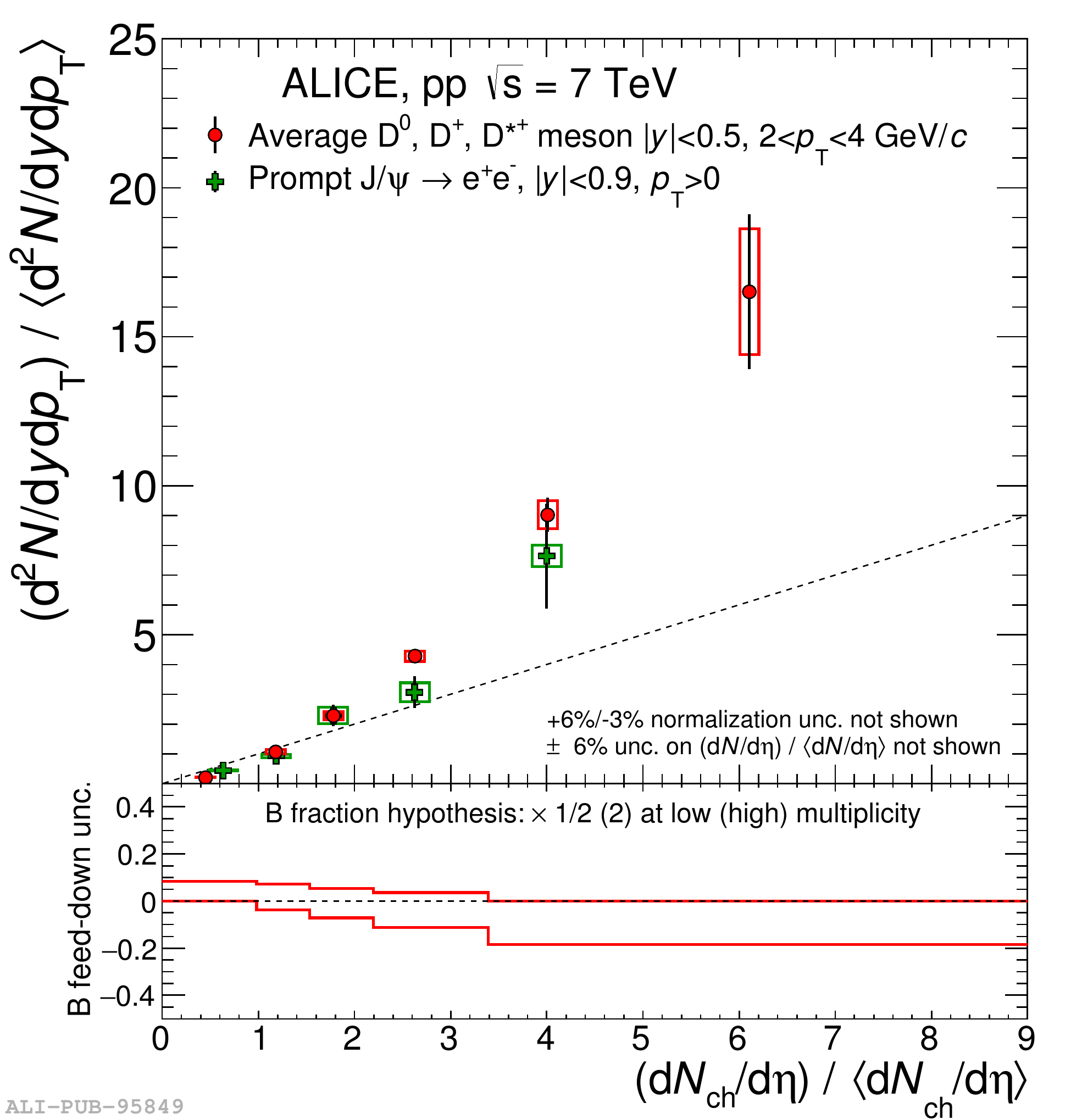}%
\caption{%
Average D-meson and inclusive J/$\psi$ relative yields as a function of the relative charged-particle multiplicity at central rapidity in pp collisions at $\sqrt{s}=7$ TeV~\cite{Adam:2015ota}.}
\label{fig:chmult}
\end{minipage}%
\hspace{0.04\textwidth}
\begin{minipage}{.48\textwidth}%
\includegraphics[width=\textwidth]{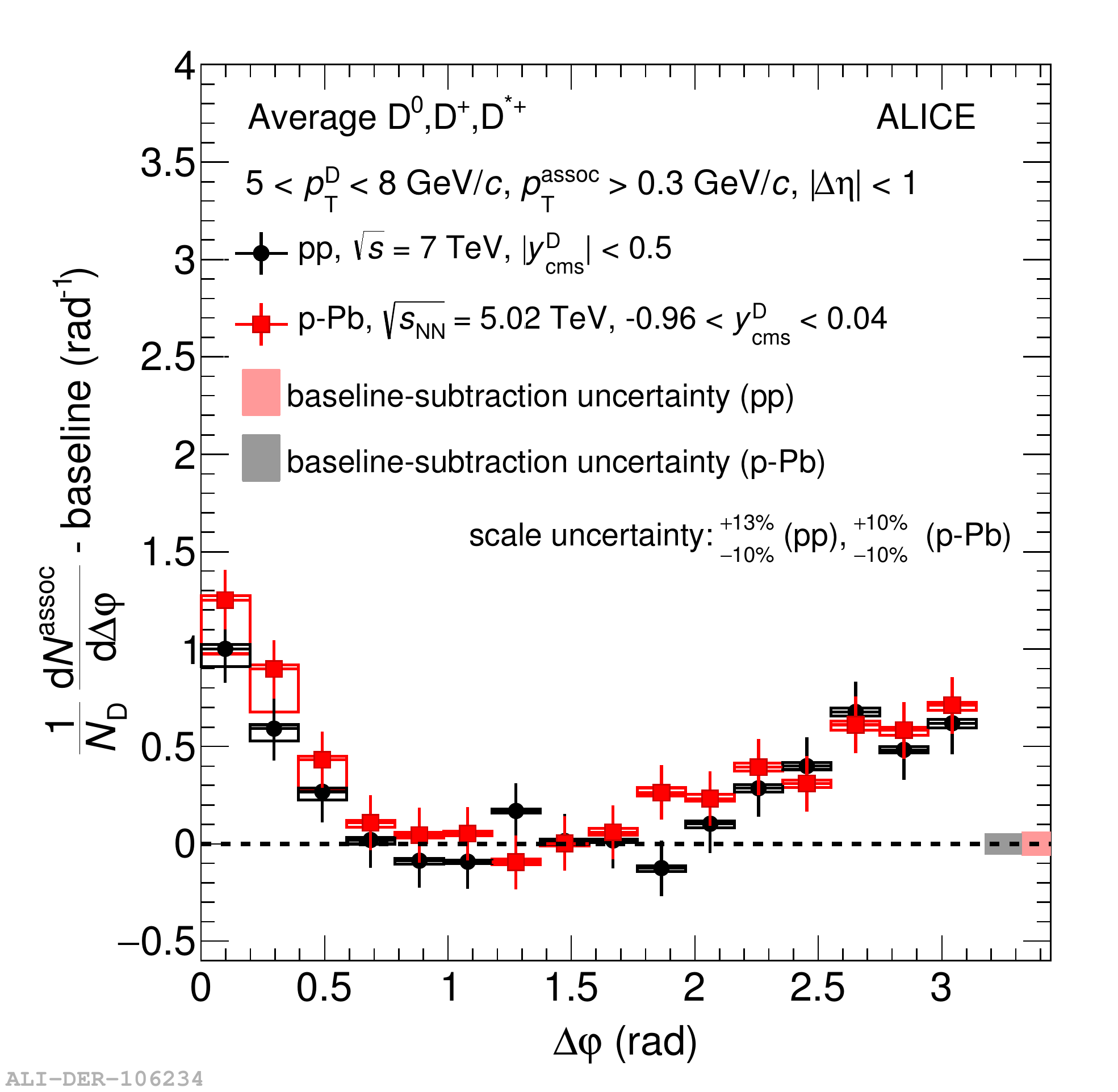}%
\caption{Baseline-subtracted azimuthal correlations of D mesons ($5<\pT<8$ GeV/$c$) and charged hadrons ($\pT>0.3$ GeV/$c$) in pp collisions at $\sqrt{s}=7$ TeV, compared to p--Pb collisions at $\sqrt{s_{\rm NN}}=5.02$ TeV~\cite{ALICE:2016clc}.}
\label{fig:corr}
\end{minipage}
\end{figure}

The average D-meson relative yield in pp collisions at $\sqs=7$ TeV exhibits a steeper-than-linear dependence on the relative charged multiplicity (a measure of event activity)~\cite{Adam:2015ota}. Figure~\ref{fig:chmult} demonstrates that the D-meson and J/$\psi$ measurements, representing open and hidden charm production, are consistent within uncertainties. 
The relative charged multiplicity of J/$\psi$ originating from B hadron decays is also consistent with that of the D mesons~\cite{Adam:2015ota}. This suggests that the multiplicity dependence is rooted in the initial hard QCD processes and is rather insensitive to hadronization.
Moreover, a similar observation was made in the multiplicity-dependent J/$\psi$ measurements by the STAR collaboration at much lower centre-of-mass (c.m.s.)\ energies of $\sqs=500$ GeV~\cite{Trzeciak:2016xyl}.
This rather universal trend can be qualitatively understood by calculations that include multiple parton interactions (MPI)~\cite{Ferreiro:2012fb}.

\section{D-h azimuthal correlations}

Further insight may be gained into charm fragmentation properties by correlation measurements~\cite{Mangano:1991jk}. Azimuthal correlations of D mesons with charged hadrons are shown in Fig.~\ref{fig:corr} after baseline subtraction. There is no difference within uncertainties between pp and p--Pb collisions, despite the different c.m.s. energies~\cite{ALICE:2016clc}. The results are consistent with different PYTHIA tunes~\cite{Sjostrand:2006za}. This attests to the observation that CNM effects at mid-rapidity are not strong enough to be visible at the current precision of our measurements~\cite{Abelev:2014hha}.

\section{Summary and outlook}

In this paper we gave an overview of selected heavy-flavour results from ALICE in pp collisions at $\sqs=7$ TeV and in p--Pb collisions at $\sqsn=5.02$ TeV. FONLL pQCD calculations describe the data from pp collisions within uncertainties. We found that nuclear modification in p--Pb collisions is generally not observable or moderate. A hint of anti-shadowing of the production of muons from heavy-flavour decays in the backward direction can be seen at low \pT{}. Looking at the production versus event activity in pp collisions one sees a universal steeper-than-linear trend that can be qualitatively explained by models with multi-parton interactions.
Azimuthal correlations of D mesons with charged hadrons show no significant difference between pp and p--Pb collisions. 

Ongoing and future measurements of the LHC Run-2 phase will extend the measurements in c.m.s. energies up to $\sqs=13$ TeV in pp, and $\sqsn=8.16$ TeV in p--Pb collisions. Measurements of jets containing a heavy ($c$ or $b$) quark will make studies of heavy-flavour fragmentation possible. The foreseen detector upgrades~\cite{Abelevetal:2014cna} for Run 3 (including the ITS, TPC and the muon forward tracker) will open the window for e.g. precision beauty measurements by a two orders-of-magnitude increase of statistics compared to the combined Run 1 and Run 2 data.


This work has been supported by the Hungarian NKFIH/OTKA K 120660 grant and the J\'anos Bolyai scholarship of the Hungarian Academy of Sciences.

\end{document}